\documentclass[a4paper, amsmath, amssymb, aps, pra, reprint, superscriptaddress, 
	showkeys, showpacs, floatfix,
	longbibliography]{revtex4-1}

\usepackage[english]{babel}
\usepackage{color}
\usepackage[pdftex]{graphicx}
\usepackage{MnSymbol}
\usepackage[center]{subfigure}
\usepackage{tabularx}
\usepackage{hyperref}
\usepackage[squaren]{SIunits}

\newcommand{\di}{\mathrm{d}}

\newcommand{\elm}{m_\mathrm{e}} 
\newcommand{\elr}{r_\mathrm{e}} 
\newcommand{\dop}{\mathrm{\Pi}} 
\newcommand{\recoil}{r}

\let\op\hat 

\newcommand{\vop}[1]{\op{\vec{#1}}} 
\let\vec\boldsymbol

\let\uvec\vec 

\newcommand{\mat}[1]{#1}
\newcommand{\dya}[2]{|#1\rangle \langle #2 | }
\newcommand{\ket}[1]{| #1 \rangle}
\newcommand{\scp}[2]{\langle #1 \hspace{0.1em}|\hspace{0.1em} #2 \rangle}
\newcommand{\exv}[3]{\langle #1 \hspace{0.1em}|\hspace{0.1em} #2 
	\hspace{0.1em}|\hspace{0.1em} #3 \rangle}
\newcommand{\mean}[1]{\langle #1 \rangle}

\begin{document}
	
	\title{Compton scattering of twisted light: angular distribution and 
		polarization of scattered photons}
	
	\author{S.~Stock}
	\email{sebastian.stock@uni-jena.de}
	\affiliation{Helmholtz-Institut Jena, Fr\"obelstieg 3, 07743 Jena, Germany}
	\affiliation{Friedrich-Schiller-Universit\"at Jena, 
		Theoretisch-Physikalisches Institut, 07743 Jena, Germany}

	\author{A.~Surzhykov}	
	\affiliation{Helmholtz-Institut Jena, Fr\"obelstieg 3, 07743 Jena, Germany}
	
	\author{S.~Fritzsche}
	\affiliation{Helmholtz-Institut Jena, Fr\"obelstieg 3, 07743 Jena, Germany}
	\affiliation{Friedrich-Schiller-Universit\"at Jena, 
		Theoretisch-Physikalisches Institut, 07743 Jena, Germany}
		
	\author{D.~Seipt}
	\email{d.seipt@gsi.de}
	\affiliation{Helmholtz-Institut Jena, Fr\"obelstieg 3, 07743 Jena, Germany}

	
	\begin{abstract}
		Compton scattering of twisted photons is investigated within a non-relativistic framework
		using first-order perturbation theory.
		We formulate the problem in the density matrix theory, which enables one to gain
		new insights into scattering processes of twisted particles by
		exploiting the symmetries of the system.
		In particular, we analyze how the
		angular distribution and 
		polarization of the scattered photons
		are affected by the
		parameters of the initial beam such as the opening angle and
		the projection of orbital angular momentum.
		We present analytical and numerical results for the angular distribution and the polarization
		of Compton scattered photons for initially twisted light and compare them
		with the standard case of plane-wave light.
	\end{abstract}
	
	\maketitle

	\section{Introduction}
	
	The inelastic scattering of photons on (quasi-)free charged particles, known also
	as the Compton scattering of light, is one of the best studied processes in quantum mechanics.
	This process demonstrates that light is more than a classical wave phenomenon and that
	quantum theory is required in order to explain the frequency shifts as well as the angular and
	polarization distribution of the scattered light \cite{Compton:PR1923,Klein:ZPhys1929}.
	For example, the classical electro-magnetic
	theory cannot properly describe the frequency shifts at low intensity of the incident photons \cite{book:Jackson},
	although these shifts are derived quite easily from the conservation of the (total) energy
	and momentum of the quantum particles involved in the scattering. In the framework of quantum
	theory, a shift in the wavelength of the photons occurs since, for an electron at rest, for example,
	a part of the incident photon energy is transferred to the recoil of the electron.
	Therefore, an elastic scattering of the light can only be assumed if the photon
	energy is negligible, compared to the rest energy of the electrons. This low-energy
	limit to the Compton scattering of photons can be described by the non-relativistic
	Schr\"o{}dinger theory \cite{book:Greiner4A}.

	Indeed, derivations of the angular distribution and polarization of the Compton
	scattered light can be found in many texts but are made usually for plane-wave photons
	and electrons \cite{book:Greiner4A,book:Landau4}.
	Such a plane-wave approximation
	to the Compton scattering applies if the lateral sizes of the incident electron and
	photon beams are (much) larger than their wavelength. In contrast, less attention has
	been paid to the Compton scattering of ``twisted" beams in which each photon (or electron)
	carries an non-zero projection of the orbital angular momentum (OAM) along the propagation direction,
	in addition to the spin angular momentum that is related to the polarization of the light
	\cite{Allen:PRA1992,ONeil:PRL2002,Molina:NatPhys2007,Yao:AdvOP2011,book:Andrews}.
	Here, we therefore investigate how an OAM of the incident radiation affects the
	angular distribution and polarization of the (Compton) scattered light. In particular,
	we apply the density matrix formalism \cite{book:Blum}
	to explore the angular distribution and polarization of the scattered light and compare this to the
	results of a plane-wave scattering. Calculations of the angle-differential Compton cross
	sections have been performed for the scattering of Bessel beams with different opening angles
	and total angular momenta $m$. While the cross section does not depend on $m$, it is highly
	sensitive with regard to the opening angle $\theta_k$ of the Bessel beams. Moreover, the formulation
	of the problem within the density matrix theory clearly highlights
	why the results do not depend on $m$.
	This can be explained as restriction to those elements of the twisted state photonic
	density matrix that are diagonal in the momentum quantum numbers
	due to the spatial symmetries of the system.

	This work is structured as follows. In Sec.~\ref{sec:theory}, we briefly review the
	non-relativistic description of Compton scattering in the framework of the density
	matrix theory.
	This includes a short account on the electron-photon interaction
	in the non-relativistic framework as well as the quantization of the photon field.
	We derive the density matrix of the Compton scattered photons
	in terms of the usual plane-wave matrix elements which are weighted
	by the initial photonic density matrix in a plane-wave basis.
	The standard case of plane-wave Compton scattering is presented for later reference in
	Sec.~\ref{sec:compton-plane}.
	The description of twisted photon states is presented in Sec.~\ref{sec:compton-twisted}.
	Detailed calculations have been performed for the angular distribution and polarization of
	Compton scattered twisted photons, and will be explained 
	in comparison with the plane-wave case.
	A short summary is finally given in Sec.~\ref{sec:conclusions}.
	In the Appendix, in addition, we collect all issues related to the normalization of
	the plane-wave and twisted-wave
	one-particle states and their density matrices.
				
	\vfill

	\section{Density Matrix Theory for non-relativistic Compton scattering}\label{sec:theory}

	We consider the scattering of plane-wave or twisted
	photons on either a beam of free electrons as emitted from an
	electron gun \cite{Englert:PRA1983}, or on a target material with a low
	work function, much lower than the frequency of light, such that the electrons
	can be considered as quasi-free.
	Twisted light has been produced in a wide range of frequencies
	\cite{Cojoc:Microelec2006,Yao:AdvOP2011,Zuerch:NatPhys2012,
	Hemsing:NatPhys2013,Bahrdt:PRL2013,Gariepy:PRL2014}.
	If the frequency of the photon $\omega$ in the
	rest frame of the electron is much smaller than the electron rest energy, i.e.~if the recoil parameter
	\begin{align}
	 \recoil = \frac{ \hbar\omega}{\elm c^2}\ll 1 \,,
	\label{eq:recoil}
	\end{align}
	we can work in the low-energy limit of non-relativistic Compton scattering.
	We then conveniently work in the rest frame of the incident electron, where the theoretical
	description of the scattering process is much easier.
	For an electron beam target
	the results observed in the laboratory frame where the electrons are moving
	can be obtained by just performing a proper Lorentz transformation.
	For electron beams with low kinetic energy,
	$E_{\rm kin} \lesssim \unit{1}{\kilo\electronvolt}$,
	the quantitative results in the rest frame and the laboratory frame differ very little.
	The size of the
	target should be larger than the lateral size of the
	vortex light beam \cite{Arlt:OptCommun2000,Kumar:OptLasEng2010,book:Andrews}.
	Throughout the paper we use units with $\hbar=1$ unless stated otherwise.
	Moreover, we employ Gaussian units, where the fine-structure constant $\alpha = e^2$.

	\subsection{Density Matrix Formalism}

	To describe the angular distribution and polarization of the
	scattered photons it is most convenient to use the density-matrix
	theory \cite{book:Blum,book:Balashov}.
	The density matrix formalism has been applied just recently to describe the
	interaction of twisted light with many-electron atoms and ions \cite{Surzhykov:PRA2015}.
	In this formalism, the system after the collision is described by the final state density
	operator $\op{\rho}_f$,
	which is related to the density operator of the system in the initial state before the scattering,
	$\op{\rho}_i$, by the scattering operator,
	\begin{align}
	\op{\rho}_f = \op{ S } \op{\rho}_i \op{  S }^\dagger \,,
	\end{align}		
	and where $\op{S}$ characterizes the interaction of the particles
	during the collision.
	
	Before the scattering,
	the electrons and photons are initially independent and uncorrelated.
	The density operator of the initial state can thus be written as the direct product of the
	electronic ($\op{\rho}_i^\mathrm{el}$) and
	photonic ($\op{\rho}_i^\mathrm{ph}$) operators \cite{Surzhykov:PRA2015}
	\begin{align}
	 \op{\rho}_i = \op{\rho}_i^\mathrm{el} \otimes \op{\rho}_i^\mathrm{ph} \,.
	\end{align}
	We describe the initial electron as a plane-wave
	in a pure quantum state $\ket{\vec p_i}$
	with the density operator
	$\op{\rho}_i^\mathrm{el} = \dya{\vec p_i}{\vec p_i}$.
	The initial photon is described by the
	initial state photonic density operator
	$\op \rho_\gamma \equiv \op \rho_i^\mathrm{ph}(\gamma)$,
	where $\gamma$ refers to a set of quantum numbers to describe that state.
	Below, we will specify the quantum numbers $\gamma$ that are needed
	to represent either a plane-wave or a twisted-wave photon.

	Let us now write the final state density operator $\op \rho_f$
	in a matrix representation in a plane-wave basis, where
	$\ket{f}=\ket{\vec p_f, \vec k_f \Lambda_f}$ abbreviates
	the final plane-wave electron
	and photon states,
	where $\vec p_f$ denotes the final electron momentum while
	$\vec k_f$ and $\Lambda_f$ stand for the final photon momentum and helicity, respectively.
	Let us also introduce complete sets of initial plane-wave states
	$\ket{i} = \ket{\vec p,\vec k \lambda}$,
	with $\sumint \limits_i \dya{i}{i} = 1$
	(for a detailed discussion of the orthonormality and completeness of these
	plane-wave bases see the Appendix),
	to obtain the density matrix in the following form:
	\begin{multline}
	\exv{f }{\op{\rho_f} }{f' } = \\
	\sumint \limits_{\lambda,\lambda'} \widetilde{\di \vec k}\widetilde{\di \vec k}'
	 \, \exv{\vec k \lambda }{\op \rho_\gamma}{\vec k' \lambda'} 
	 \exv{ \vec p_i, \vec k' \Lambda'  }{\op S^\dagger }{f'} 
	 \exv{f}{\op S}{ \vec p_i, \vec k \Lambda } \,,
	\label{eq:density.matrix}
	\end{multline}		
	where we employed the fact that the initial electron is in a plane wave state as discussed above,
	and where $\widetilde{\di \vec k}$ is an abbreviation for the properly normalized integration
	measure (for details we refer to the Appendix). 
	This equation states that in order to calculate the final state density matrix for
	an arbitrary initial photon state $\op \rho_\gamma$, either plane-wave or a twisted-wave or
	any other photon state, we just need to know the ordinary plane-wave $S$ matrix elements
	to describe the physics. Equation \eqref{eq:density.matrix} describes how these
	plane-wave $S$ matrix elements have to be weighted by the elements of the initial photonic density
	matrix in the plane-wave basis $\exv{\vec k \lambda }{\op \rho_\gamma}{\vec k' \lambda'} $.

	The elements of the $S$ matrix themselves can be represented in a form
	\begin{align}
	\exv{f}{\op S}{i} = -2\pi i \delta(E_i - E_f)  \exv{f}{\op T}{i} \,,
	\label{eq:Smatrix}
	\end{align}		
	where the delta function ensures the conservation of energy, i.e.~the total energy of the initial
	state particles $E_i$ equals the total energy of the final state particles $E_f$.
	The matrix elements of the transition operator $\op{T}$ can be calculated using
	perturbation theory \cite{book:Greiner4A}.

	The normalization of the density matrix is purely conventional and for us it is most convenient to
	normalize it to the total cross section.
	This is achieved by dividing out the flux of incident particles \cite{book:Taylor}:
	\begin{align}
	\sigma := \frac{ {\rm tr} \, \op{\rho}_f }{TV {\rm tr}\, ( \op{J} \op{\rho}_i )}  \,,
	\label{eq:cross.section}
	\end{align}	
	where $T$ and $V$ are the interaction time and volume, respectively.
	In order to calculate the cross section for twisted particles, which are spatially localized,
	we need a convenient definition of flux density operator $\op J$ in terms of the
	densities of colliding particles times their relative velocity \cite{book:Landau2}.
	Since twisted beams are not spatially homogeneous perpendicular to their propagation
	direction, a proper definition of the cross section
	according to the book of Taylor \cite{book:Taylor}
	includes an average over the lateral structure of the beam of incident particles.
	We conclude that we should define the cross section by means of
	the spatially averaged density of the initial particles
	$\mean{n} := V^{-1}\int \! \di^3x \, n(\vec x)$ that is
	proportional to the trace of the initial state density operator
	$\mean{n} ={ \rm tr} \, \op \rho / V $.
	Thus, the operator of the averaged density is just the
	unity operator divided by the quantization volume $V$
	(see the Appendix), and the flux density operator $\op J$
	in \eqref{eq:cross.section} can be represented
	as $\op J = c \frac{\op{1}_\mathrm{el} }{V} \otimes \frac{ \op{1}_\mathrm{ph}}{V}$,
	i.e.~by the relative velocity, the speed of light $c$, times the operators of the averaged
	particle densities of electrons and photons in the initial state.

	Using the basis expansion of the density operator, Eq.~\eqref{eq:density.matrix},
	and by employing the energy conservation in \eqref{eq:Smatrix}
	we find that the cross section \eqref{eq:cross.section} contains a factor
	$\delta(E_f-E_i) \delta(E_f-E_{i'}) = \delta(E_f-E_i) \delta(E_i-E_{i'})$.
	For a non-zero contribution to the scattering
	cross section, the total energies of the initial state bases used for the expansion
	of the final state density matrix, Eq.~\eqref{eq:density.matrix}, need to be equal.
	Finally, we can express the scattering cross section as
	\footnote{By using the relation for the energy delta
	function
	$\lim_{E_i'\to E_i} 2\pi \delta(E_i-E_i')
	=
	 \lim_{E_i'\to E_i} \intop \limits_{-T/2}^{T/2} \! \di t \, e^{i (E_i-E_i' )t} 
	= T$ in the limit $T\to \infty$ \cite{book:Peskin}.
	}
	\begin{multline}
	\sigma = \frac{ 2\pi V }{ c\, {\rm tr}\, \op{\rho}_\gamma  } 
	\sumint \limits_f \sumint \limits_{\lambda, \lambda'}
	 \widetilde{\di \vec k} 
	\widetilde{\di \vec k}'	
	\delta(E_i-E_f)
	\\
	\times
		 \exv{\vec k \lambda }{\op \rho_\gamma}{\vec k' \lambda'} 
	 \exv{\vec p_i, \vec k'\lambda'}{\op T^\dagger }{f} 
	 \exv{f}{\op T}{\vec p_i, \vec k \lambda } \,.
	 \label{eq:cross.section.final}
	\end{multline}
	Here we already employed
	the normalization of the initial electron plane-wave states
	${\rm tr}\, \op \rho_{i}^\mathrm{el}= \scp{\vec p_i}{\vec p_i} = 1$.
	This concludes our general discussion of the density matrix formalism
	and the definition of the cross section. The main result of this section
	is the representation of the final-state density matrix \eqref{eq:density.matrix}
	of Compton scattered light for an incident plane-wave or twisted-wave
	photon beam---characterized by the photonic density operator
	$\op \rho_\gamma$---in terms of
	the plane-wave matrix elements for the interaction of electrons and photons.

	\subsection{Interaction Between Photons and Electrons}

	Let us now turn our attention on the description of the interaction between
	the electrons and the photons.
	The form of the interaction Hamiltonian,
	\begin{align}
	\op{H}_\mathrm{int} = \frac{e \vop{A} \cdot \vop p }{\elm c} + 
	\frac{e^2 \vop{A}^2  }{2\elm c^2}\,,
	\label{eq:H.int}
	\end{align}
	follows from the gauge invariant minimal
	coupling of the electromagnetic field to the free electron Hamiltonian \cite{book:Greiner4A}.
	The operator  $\vop A$ of the electromagnetic vector potential
	describes the emission or the absorption of one photon \cite{book:Greiner4A}.
	Compton scattering is a two-photon process: The incident photon is absorbed by
	the electron while the scattered photon is emitted into some other direction.
	The one-photon interaction operator $\vop A \cdot \vop p$ does not
	contribute to the non-relativistic Compton scattering amplitude, because
	the matrix elements of the electron momentum
	operator $\op p$ vanish in the rest frame of the electron \cite{book:Greiner4A}.
	Thus, the Compton scattering amplitude can be calculated in first-order perturbation
	theory by means of the two-photon contribution  $\vop A^2$ to the Hamiltonian;
	the transition matrix elements are just
	given by $\exv{f}{\op{T}} {i} = \exv{f}{\op{H}_\mathrm{int}}{i}$.

	The photon field operator $\vop A$ that enters the interaction Hamiltonian, Eq.~\eqref{eq:H.int},
	can be represented by
	its mode expansion into a circularly polarized plane-wave basis
	$\vec u_{\vec k \Lambda}(\vec x) 
		= 
		e^{i\vec k\cdot \vec x} \, \vec{\varepsilon}_{\vec k \Lambda}$,
	where the polarization vector $\vec{\varepsilon}_{\vec k \Lambda}$
	is perpendicular to the wave-vector,
	$\vec k \cdot \vec{\varepsilon}_{\vec k \Lambda}=0$.
	There are two independent solutions for each $\vec k$,
	denoted by the photon helicity $\Lambda = \pm1$.
	In terms of these plane-wave modes the photon field operator is given by
	\begin{align}
		\vop{A}(\vec{x}) 
		= \sumint \limits_\Lambda \! \widetilde{\di \vec{k}}  \, N_k
			\left[
				\op{c}_{\vec k \Lambda} \vec u_{\vec k \Lambda}(\vec x)
				+ 
				\op{c}_{ \vec k \Lambda}^\dagger \vec u_{\vec k \Lambda}^* (\vec x)
			\right] \,,
		\label{eq:vector-potential}
	\end{align}
	where we employ the proper integration measure $\widetilde{\di \vec{k}}$
	to ``count'' the basis functions
	(see the Appendix for details).
	The creation operator $\op{c}^\dagger_{\vec k\Lambda}$
	creates a normalized one-photon plane-wave state from the vacuum
	$\ket{\vec k \Lambda } = \op{c}^\dagger_{\vec k\Lambda} \ket 0$ that is characterized
	by its linear momentum (wave-vector) $\vec k$ and helicity $\Lambda$.
	The one-photon states are normalized as $\scp{\vec k\Lambda}{\vec k \Lambda} = 1$.
	Moreover, the normalization factor $N_k = \sqrt{ 2\pi c / k V }$ is determined such that
	the energy eigenvalues of the free field Hamiltonian are just $\omega_k =  c |\vec k|$
	for one-photon states $\ket{\vec k\Lambda}$ \cite{book:Greiner4A}.
	It is sufficient to know the above representation of the photon field operator in a plane-wave basis
	because we just need to calculate the plane-wave matrix elements
	by means of Eqs.~\eqref{eq:density.matrix} and \eqref{eq:cross.section.final}.

	\subsection{The Reduced Density Matrix of Compton Scattered Photons}

 	We are going to investigate the angular distribution and polarization of the scattered photons,
 	and we are not interested in the electron distribution after the scattering has occurred.
 	We therefore have to calculate a reduced density matrix
 	$\rho_{\Lambda' \Lambda}(\uvec k_f/k_f)$,
 	which depends on the direction of the scattered photon $\uvec k_f/k_f$
 	and its polarization state $\Lambda$,
 	by tracing out the unobserved final electron states.
	We obtain the reduced density matrix of the Compton scattered photons
	\begin{widetext}
	\begin{align}
	\rho_{\Lambda' \Lambda} (\uvec k_f/k_f) 
	= 
	\frac{V^2}{(2\pi)^2 c \,  {\rm tr} \, {\op \rho_\gamma } }
	  \sumint \limits_{\lambda,\lambda'} \! \widetilde{\di \vec k} \widetilde{\di \vec k}' \,
	  \int \! \widetilde{\di \vec p_f }  \, 
	  \int \! \di k_f k_f^2 \, \delta( E_f - E_i ) \,
	  \exv{\vec k\lambda}{\op \rho_\gamma }{\vec k'\lambda'} \,
	 \mathcal  M^*_{\vec k' \lambda'}(\Lambda')
	  \mathcal M_{\vec k \lambda}(\Lambda)  \,,
	 \label{eq:def.reduced.density.matrix2} 
	\end{align}
	\end{widetext}
	in terms of the plane-wave matrix elements
	$\mathcal M_{\vec k \lambda}(\Lambda)
	 = \exv{\vec p_f; \vec k_f \Lambda}{ \op H_\mathrm{int}}{ \vec p_i; \vec k \lambda }$.
	The calculation of the plane-wave matrix elements for non-relativistic Compton scattering
	can be found in textbooks, e.g.~in \cite{book:Greiner4A}, and we only cite here the final result:
	\begin{align}
	\mathcal M_{\vec k \lambda}(\Lambda)
	 &=  \frac{e^2}{\elm c^2} \frac{2\pi  c^2}{L^3} \,
			\frac{\vec{\varepsilon}_{\vec{k}_f\Lambda}^* \cdot \vec{\varepsilon}_{\vec{k} \lambda} }
			{\sqrt{\omega_i \omega_f} } \, 
			\widetilde{\delta} (\vec{p}_i + \vec{k} - \vec{k}_f-\vec{p}_f)\,.
			\label{eq:Mpws}
	\end{align}
	Here, $\tilde \delta$ denotes a normalized delta function with the property
	$\tilde \delta(0)=1$. For details we refer to the Appendix.
	Because the reduced density matrix \eqref{eq:def.reduced.density.matrix2} contains the product
	of two plane-wave matrix elements we also get two delta functions
	that ensure the conservation of momentum. Their product can be reformulated as,
	\begin{multline}
	\widetilde{\delta} (\vec p_i +\vec k - \vec p_f - \vec k_f) 	
	\widetilde{\delta} (\vec p_i +\vec k' - \vec p_f - \vec k_f) \\
	= 
	\widetilde{\delta} (\vec k'-\vec k) 
	\widetilde{\delta} (\vec p_f +\vec k_f - \vec p_i - \vec k) \,, 
	\end{multline}		
	where we obtain a factor $\widetilde{\delta} (\vec k'-\vec k)$, which consumes one of the
	integrations over the plane-wave bases, $\widetilde{\di \vec k'}$,
	in Eq.~\eqref{eq:def.reduced.density.matrix2}.
	For this reason, only those elements of the
	initial state photonic density matrix with $\vec k=\vec k'$ remain
	in Eq.~\eqref{eq:def.reduced.density.matrix2}.
	 Thus, just the \textit{momentum-diagonal elements} of $\op \rho_\gamma$
	 contribute to the reduced density matrix
	of the scattered photons
	 and the coherences in the
	off-diagonal elements of the initial photonic density matrix are lost.
	The reason for this behaviour is of course the momentum conservation
	in the plane-wave matrix elements that is related
	to the spatial homogeneity of the system via Noether's theorem \cite{Noether:NachGesWissGoett1918}.
	All particles are described as plane-waves, except for the initial photons
	that are prepared in the hitherto unspecified quantum state $\op \rho_\gamma$.
	It is the spatial homogeneity of the residual system of incident and scattered
	particles which excludes the interference of different momentum
	components of the initial photonic state
	$\op \rho_\gamma$ from the
	reduced density matrix \eqref{eq:def.reduced.density.matrix2} of the scattered photons.

	The above momentum conservation,	
	together with the conservation of the total energy
	$\delta(E_f - E_i) = \delta(\omega_f + p_f^2/2\elm - \omega - p_i^2/2\elm)$,
	determines the frequency of the scattered photons.
	Recalling that we work in the rest frame of the incident
	electron, $\vec p_i=0$, we find
	for the frequency of the scattered photons
	\begin{align}
		\omega_f = \omega \left[  1 - \frac{\hbar \omega}{\elm c^2} (1 - \cos \theta )
		{+ O \left( \frac{\hbar\omega}{\elm c^2} \right)^2 }\right] \,,
		\label{eq:omegaf}
	\end{align}
	where we temporarily reinstated $\hbar$.
	The expression for $\omega_f$ in \eqref{eq:omegaf} accounts for the well-known frequency red-shift
	of Compton scattering \cite{Compton:PR1923}, which depends on the angle $\theta$ between
	the initial and the scattered photon momentum vectors.
	When starting from a fully relativistic QED calculation,
	we would get the above non-relativistic frequency shift in \eqref{eq:omegaf}
	the leading order of an expansion in the small recoil parameter $\recoil$, Eq.~\eqref{eq:recoil}.
	Because of Eq.~\eqref{eq:omegaf}, the length of the scattered photon's wave-vector
	$|\vec k_f| = \omega_f(\theta)/c$ is completely determined by its direction.

	In the electric dipole approximation, the momentum of the photon
	and its recoiling effect on the electron
	is neglected, $\vec k=\vec k_f=0$.
	This is a good approximation whenever the recoil parameter,
	Eq.~\eqref{eq:recoil}, is negligibly small.
	The non-relativistic Compton scattering
	becomes elastic within the dipole approximation: $\omega_f=\omega$.
	Moreover, the dipole approximation coincides with the formal classical limit $\hbar \to 0$.
	Within the dipole approximation we obtain as final formula for the elements of the reduced density matrix
	the following expression:
	\begin{widetext}
		\begin{align}
			\rho_{\Lambda' \Lambda} (\uvec k_f/k_f) 
			= 
			\frac{e^4}{\elm^2 c^4  }
			\sumint \limits_{\lambda,\lambda'} \! \widetilde{\di \vec k} \, \,
			\frac{ \exv{\vec k \lambda }{\op \rho_\gamma }{\vec k\lambda'} }{  {\rm tr} \,
				{\op \rho_\gamma }  } \,
			\left(
						\vec{\varepsilon}_{\vec{k}_f \Lambda'}^*\cdot
						\vec{\varepsilon}_{\vec{k} \lambda'}
				\right)^*
			\left( 	
						\vec{\varepsilon}_{\vec{k}_f\Lambda}^*\cdot
						\vec{\varepsilon}_{\vec{k}\lambda}
			\right) \,.
			\label{eq:reduced.density.matrix.finals}
		\end{align}
	\end{widetext}

	\subsection{Stokes parameters and differential cross section}

	In the previous subsection we calculated a suitable expression for the reduced density matrix of
	Compton scattered photons, Eq.~\eqref{eq:reduced.density.matrix.finals}.
	We now relate the elements of the reduced density matrix to the angular distribution
	of the scattered photons, i.e.~to their angular differential cross section and the polarization properties.
	According to \cite{book:Blum,book:Balashov}, the reduced density matrix 
	can be represented by the three Stokes parameters
	$\vec P = (P_1,P_2,P_3)$ via 
	\begin{align}
		\mat{\rho}_{\Lambda'\Lambda} (\uvec k_f) =  \frac{\di \sigma}{\di \Omega}
		\,
		\frac{1}{2}
			\begin{pmatrix}
				1+P_3 & P_1- i P_2 \\ 
				P_1+ i P_2 & 1 - P_3
			\end{pmatrix}_{\Lambda'\Lambda}\,.
			\label{eq:reduced.density.matrix.stokes}
	\end{align}
	From this representation it is easy to obtain the angular differential
	cross section of Compton scattered photons
	as
	\begin{align}
	 \frac{\di \sigma}{\di \Omega} = \sum_{\Lambda=\pm1} \rho_{\Lambda\Lambda}(\uvec k_f)  
	 = \rho_{+1+1} + \rho_{-1-1} \,.
	 \label{eq:cs.sum.rho}
	\end{align}
	The Stokes parameters are given by
	\begin{align}
	P_1 &= \frac{\rho_{+1-1} + \rho_{-1+1}}{\rho_{+1+1}+\rho_{-1-1}} \,,\\
	P_2 &= \frac{i\rho_{+1-1} - i\rho_{-1+1}}{\rho_{+1+1}+\rho_{-1-1}}	\,,\\
	P_3 &= \frac{\rho_{+1+1} - \rho_{-1-1}}{\rho_{+1+1}+\rho_{-1-1}} \,.
	\end{align}
	As usual, the Stokes parameters $P_1$ and $P_2$ represent the intensity of
	light that is linearly polarized under different angles with respect to the scattering plane.
	The scattering plane is defined by the direction of the incident beam of light and by
	the momentum vector of the scattered photon.
	The Stokes parameter $P_3$ measures the amount of light with circular polarization 
	\cite{book:Balashov,McMaster:RevModPhys1961}.
	Moreover, the degree of polarization $\dop$ is defined as the length of the vector $\vec P$
	\begin{align}
		\dop = \sqrt{P_1^2+P_2^2+P_3^2} \,.
	\end{align}
	This concludes our discussion of the density matrix formalism. We are now ready to study the
	angular distribution and the polarization properties of Compton scattered light for both
	plane-wave and twisted photons.

	\section{Compton Scattering of Plane-wave Photons}\label{sec:compton-plane}

	Let us now apply the formalism to the standard case of the Compton scattering of plane-wave photons,
	as a starting point for later comparison with the case of twisted light.
	Moreover, this will convince us that we normalized the reduced density matrix correctly
	to obtain the differential cross section by comparing with the well known results from the literature.

	We now specify the initial photon state as a plane wave with
	wave-vector $\vec k_i$ and in a well defined helicity state $\Lambda_i$,
	with the photonic initial density operator
	$\op \rho_\gamma = \op \rho_{\vec k_i \Lambda_i} = \dya{\vec k_i \Lambda_i}{\vec k_i \Lambda_i}$.
	It has the following representation in a plane-wave basis:
	\begin{align}
		\exv{\vec k \lambda }{\op \rho_{\vec k_i \Lambda_i}}{ \vec k' \lambda'} 
		=	
		\widetilde{\delta} (\vec k-\vec k')  \delta_{\lambda\lambda'}  
		\widetilde{\delta} (\vec k-\vec k_i) \delta_{\lambda\Lambda_i} \,.
	\end{align}
	Using its diagonal elements
	$\exv{\vec k \lambda }{\op \rho_{\vec k_i \Lambda_i}}{ \vec k \lambda'} 
		=	
		\delta_{\lambda\lambda'}  \delta_{\lambda\Lambda_i} 
		 \widetilde{\delta} (\vec k-\vec k_i)$
	into Eq.~\eqref{eq:reduced.density.matrix.finals}
	readily yields the reduced density matrix of non-relativistic Compton scattering
	of circularly polarized plane-wave photons in the dipole approximation
	\begin{align}
		\rho_{\Lambda'\Lambda} (\uvec k_f ) &= 
		\elr^2
		\left(
					\vec{\varepsilon}_{\vec{k}_f \Lambda'}^*\cdot
					\vec{\varepsilon}_{\vec{k}_i \Lambda_i}
			\right)^*
		\left( 	
					\vec{\varepsilon}_{\vec{k}_f\Lambda}^*\cdot
					\vec{\varepsilon}_{\vec{k}_i\Lambda_i}
		\right) \,,
		\label{eq:reduced.density.pw}
	\end{align}		
	where $\elr=e^2/(\elm c^2)\simeq \unit{2.8}{\femto\metre}$ is the classical electron radius.

	To become more specific, we need to specify the scattering geometry, and to express the
	photon polarization vectors in terms of the scattering angles.
	Let us assume that the initial photon beam propagates along the $z$-axis, $\vec k_i/k_i=(0,0,1)^T$,
	while the scattered photon propagates into the direction
	$\vec{k}_f /k_f = (\sin\theta \cos\varphi,\,
							     \sin\theta \sin\varphi, \,
							      \cos \theta)^T$,
	where the scattering angle $\theta$
	and azimuthal angle $\varphi$
	are the usual polar and azimuthal angles in spherical coordinates.
	For the polarization vectors of the scattered photons we give the explicit representation
	\begin{align}
		\vec{\varepsilon}_{\vec{k}_f \Lambda} =
		\vec{\varepsilon}_\Lambda(\theta,\varphi) = 
		\frac{1}{\sqrt{2}}
		\begin{pmatrix}
			\cos\theta\cos\varphi- i \Lambda\sin\varphi \\
			\cos\theta\sin\varphi + i \Lambda\cos\varphi \\
			-\sin\theta
		\end{pmatrix}\,, 
		\label{eq:def.polvector}
	\end{align}
	which makes evident that the polarization vector is orthonormalized
	$ \vec{\varepsilon}_{\vec{k}_f \Lambda}^* \cdot \vec{\varepsilon}_{\vec{k}_f \Lambda'}
	= \delta_{\Lambda \Lambda'}$
	and perpendicular to the momentum direction
	$\vec{k}_f \cdot \vec{\varepsilon}_{\vec{k}_f \Lambda} = 0$.
	Because the incident photon propagates along the $z$-axis,
	its polarization vector is just 
	$\vec{\varepsilon}_{\vec k_i \Lambda_i} 
				= \vec{\varepsilon}_{\Lambda_i}(0,0) 
				= ( 1/\sqrt{2} ,  i\Lambda_i/\sqrt{2}  , 0)^T$.

	The differential Compton cross section as a function of the scattering angle
	is just given by using the representations \eqref{eq:def.polvector} of the photons polarization
	vectors into Eq.~\eqref{eq:reduced.density.pw}, and
	by summing over the final state polarization according to Eq.~\eqref{eq:cs.sum.rho}.
	This yields
	\begin{align}
		\frac{\di\sigma_\mathrm{pw}}{\di\Omega} 
		= \frac{\elr^2}{2} (1+\cos^2\theta)\,,
		\label{eq:pw-cs}
	\end{align}
	where $\theta$ denotes the scattering angle, i.e.~the angle between the 
    wave-vectors of the incident and the scattered light.
	The result \eqref{eq:pw-cs} for the angular differential cross section
	of Compton scattered light in the dipole
	approximation is well known and can also be obtained by means 
	of classical electrodynamics \cite{book:Jackson}.	
	An integration of Eq.~\eqref{eq:pw-cs} over all directions of the scattered photons just
	yields the well known total Thomson cross section
	$\sigma = \frac{8}{3}\pi \elr^2 \simeq \unit{665}{\milli\barn}$.
	This comparison with well-known results from the literature
	shows that we normalized the final state density matrix correctly to the cross section.

	Similarly to the cross section, we can obtain explicit expressions
	for the three Stokes parameters of the scattered photons as
	\begin{align}
		P_1 &= -\frac{\sin^2\theta}{1+\cos^2\theta}\,, 
		\label{eq:P1.pw}\\
		P_2 &= 0\,, \\
		P_3 &= \frac{2\Lambda_i \cos\theta}{1+\cos^2\theta}\,.
		\label{eq:P3.pw}
	\end{align}
	As we see from Eqs.~\eqref{eq:P1.pw} -- \eqref{eq:P3.pw},
	the scattered photons are not necessarily circularly polarized,
	although the initial photons were.
	The ratio of the amount of linearly and circularly polarized photons varies with the 
	scattering angle $\theta$.
	For instance, under $\theta=\unit{90}{\degree}$ we have $P_1=-1$ and
	the photons are completely linearly polarized
	in the direction perpendicular to the scattering plane.
	Nevertheless, the scattered photons are fully polarized, $\dop=1$,
	independent of the scattering angle.
	This concludes our short review of the analytic results for the differential cross section and
	the Stokes parameters for plane-wave photons. We will pick them up in the next section where
	we compare them with the case Compton scattering of twisted light.

	\section{Compton Scattering of Twisted-wave Photons}\label{sec:compton-twisted}

	We now turn to the main aspect of our paper: The calculation of the angular distribution
	and the polarization properties of Compton scattered twisted light.
	For that, we need first to construct the initial photonic density matrix for twisted photons.

	\subsection{Description of Twisted Photon States and the Twisted Photon Density Matrix}

	The state of a photon in a Bessel beam that propagates along the $z$-axis,
	briefly referred to as a Bessel-state or twisted photon,
	is characterized by its longitudinal momentum $\kappa_\parallel$, i.e.~the component
	of the linear momentum along the beam's propagation axis,
	the modulus of the transverse momentum $\kappa_\perp = |\vec k_\perp|$,
	the projection of the total angular momentum (TAM) onto the propagation axis $m$,
	and the photon helicity $\Lambda$ \cite{Matula:JPB2013}.
	A photon in a twisted one-particle Bessel-state is, thus, characterized by the quantum numbers
	$\ket{\gamma} =\ket{\kappa_\perp \kappa_\parallel m \Lambda}$.
	It can be represented by a coherent superposition of plane-wave states
	\cite{Jentschura:PRL2011,Ivanov:PRD2011,Matula:JPB2013,Surzhykov:PRA2015},
	\begin{align}
	\ket{\kappa_\perp \kappa_\parallel m \Lambda} 
		=
		 \int \! \widetilde{\di \vec k} \, 
		 b_{\kappa_\perp\kappa_\parallel m} (\vec k) \, 
		 \ket{\vec k \Lambda} \,,
		 \label{eq:tw.superposition}
	\end{align}
	where we use the proper integration measure $\widetilde{\di \vec k}$ for plane-wave states
	(see the Appendix). From Eq.~\eqref{eq:tw.superposition} we see that $\Lambda$ refers
	to the helicity of the plane-wave components of the twisted photon.
	The amplitudes $b_{\kappa_\perp\kappa_\parallel m} (\vec k)$ are given by
	\begin{align}
	b_{\kappa_\perp\kappa_\parallel m} (\vec k)
	 = N_{\rm tw} \, \delta(k_z-\kappa_\parallel) 	\, 
	 		a_{\kappa_\perp m}(\vec k_\perp)
	\label{eq:tw.b.amplitude}
	\end{align}		
	and are related to the usually employed \textit{transverse} amplitudes
	(see e.g.~\cite{Jentschura:PRL2011,Matula:JPB2013,Surzhykov:PRA2015})
	\begin{align}
	a_{\kappa_\perp m}(\vec k_\perp) 
	= \sqrt{\frac{2\pi}{\kappa_\perp}} (-i)^m e^{im\varphi_k} \delta( |\vec k_\perp| -\kappa_\perp) \,.
		\label{eq:tw.a.amplitude}
	\end{align}		
	Because for a photon in a twisted Bessel state
	$\kappa_\parallel$ and $\kappa_\perp$ are well-defined, all
	the momentum vectors $\vec k$ of the superposition \eqref{eq:tw.superposition}
	are lying on a cone in momentum space with fixed opening angle
	$\theta_k = \arctan(\kappa_\perp/\kappa_\parallel)$.
	The direction of the momentum vector $\vec k$ on the cone is undefined,
	and can be parametrized as
		\begin{align}
		\vec{k} = \vec{k}(\varphi_k) =
			\begin{pmatrix} 
				\kappa_{\perp}\cos\varphi_k \\ 
				\kappa_{\perp}\sin\varphi_k \\ 
				\kappa_\parallel 
			\end{pmatrix} 
		= k \begin{pmatrix} 
				\sin\theta_k\cos\varphi_k \\ 
				\sin\theta_k\sin\varphi_k \\ 
				\cos\theta_k 
			\end{pmatrix}\,.
			\label{eq:k.cone}
	\end{align}
	where $\varphi_k$ is the azimuthal angle
	that defines the orientation of one particular
	vector $\vec k(\varphi_k)$ on the momentum cone.
	The length of these vectors,
	$k = |\vec k(\varphi_k)| = \sqrt{ \kappa_\parallel^2 + \kappa_\perp^2}$,
	are related to the photon frequency $\omega=ck$ as for plane-waves.
		
	The normalization factor $N_{\rm tw} = \sqrt{4\pi^3/L_zRV}$ that appears
	in the definition of the amplitudes
	$b_{\kappa_\perp \kappa_\parallel m \Lambda}$
	is determined such that the twisted one-particle states are
	orthonormalized in the following way:
	\begin{multline}
	\scp{\kappa_\perp \kappa_\parallel m \Lambda}{\kappa_\perp' \kappa_\parallel' m' \Lambda'} 
	\\
	= \frac{2\pi^2}{RL_z} \delta_{mm'} \delta_{\Lambda\Lambda'}
	\delta(\kappa_\perp-\kappa_\perp') 	\delta(\kappa_\parallel-\kappa_\parallel') 
	\end{multline}
	and
	$\scp{\kappa_\perp \kappa_\parallel m \Lambda}{\kappa_\perp \kappa_\parallel m \Lambda} = 1$.
	This corresponds to the normalization to one particle per cylindrical volume $V = \pi R^2 L_z$,
	where both the radius $R$ and the length $L_z$ of the cylinder are going to infinity
	(see the Appendix).

	Let us now construct the density operator for twisted photons
	in the pure quantum state \eqref{eq:tw.superposition}, together with its matrix representation
	in a plane-wave basis. The latter is needed
	to calculate the reduced density matrix, Eq.~\eqref{eq:reduced.density.matrix.finals}, of
	Compton scattered twisted light.
	The normalization of the one-photon states implies that
	the twisted-state density operator
	\begin{align}
	\op \rho_\gamma = 
	\op \rho_{\kappa_\perp \kappa_\parallel m \Lambda}
			= \dya{\kappa_\perp \kappa_\parallel m \Lambda}{\kappa_\perp \kappa_\parallel m \Lambda}
	\label{eq:rho.tw}
	 \end{align}
	has unity trace, ${\rm tr} \, 	( \op \rho_{\kappa_\perp \kappa_\parallel m \Lambda} ) = 1$,
	i.e.~it is normalized to an average particle density of ``one particle per volume $V$''.
	The matrix elements of the twisted density operator \eqref{eq:rho.tw} in a plane-wave basis are just given
	by products of the amplitudes $b_{\kappa_\perp \kappa_\parallel m \Lambda}$,
	Eq.~\eqref{eq:tw.b.amplitude},
	\begin{align}
	\exv{\vec k \lambda }{\op \rho_{\kappa_\perp\kappa_\parallel m \Lambda} }{\vec k'\lambda'} 
	&= \delta_{\lambda\lambda'} \delta_{\lambda \Lambda} \,
	b_{\kappa_\perp\kappa_\parallel m}(\vec k) \,
	b^*_{\kappa_\perp\kappa_\parallel m}(\vec k') \nonumber \\
	&\propto \delta_{\lambda\lambda'} \delta_{\lambda \Lambda}  e^{im(\varphi_k-\varphi_k')}
	\label{eq:rho_ph.tw}
	\end{align}
	and they are diagonal in the helicity quantum numbers.
	We stress that only the \textit{momentum-off-diagonal} elements of the density matrix
	\eqref{eq:rho_ph.tw} do depend on the projection of total angular momentum $m$.
	It enters as the difference of the vortex phase factors
	$e^{im\varphi_k}$ of the two plane-wave components
	$\vec k\neq \vec k'$. 
	On the other hand, the \textit{momentum-diagonal} elements of the above density matrix
	\begin{multline}
	\exv{\vec k \lambda }{\op \rho_{\kappa_\perp\kappa_\parallel m \Lambda} }{\vec k \lambda'} 
	\\
	= \delta_{\lambda\lambda'} \delta_{\lambda \Lambda} \frac{(2\pi)^2 }{ V \kappa_\perp}
	 \delta(k_z-\kappa_\parallel) \delta(|\vec k_\perp| - \kappa_\perp)\,
\label{eq:rho.twisted}
	\end{multline}
	which enter the calculation of the reduced density matrix of Compton scattered photons, 
	Eq.~\eqref{eq:reduced.density.matrix.finals}, 	
	are completely independent of $m$. Therefore, also the angular distribution and
	polarization of the Compton scattered photons do
	not dependent of $m$.

	A few remarks might be in order why the reduced density matrix of the
	scattered photons is independent from $m$.
	In this paper we are interested in the angular distribution and polarization properties of the
	Compton scattered photons.
	We, therefore, project the final state density operator onto
	a basis of plane-wave states that can be observed by a usual
	detector, which measures the linear momentum of a photon \cite{Ivanov:PRD2011}.
	Thus, all involved particles except for the initial twisted photons
	are described as plane-wave states.
	From the discussion in the previous section we know that the
	momentum conservation described by the delta function in the plane-wave
	matrix elements enforces the restriction to the momentum-diagonal
	elements of the initial photonic density matrix.
	Thus, we loose the dependence on $m$ because only
	the momentum-diagonal elements of the photonic density matrix
	\eqref{eq:rho_ph.tw} that describes the
	initial twisted state contribute due to symmetries of the system.

	It is known from previous studies that the scattering of
	twisted particles on spatially homogeneous systems, such as
	plane-waves \cite{Ivanov:PRD2011,Seipt:PRA2014b},	
	impact-parameter averaged atomic targets
	\cite{Marggraf:PRA2014,Surzhykov:PRA2015},
	impact-parameter averaged potential scattering
	\footnote{V.~G.~Serbo, I. V. Ivanov, S. Fritzsche, D. Seipt, and A. Surzhykov,
	``Scattering of twisted relativistic electrons by atoms'', submitted},
	leads to angular distributions of the scattered particles
	or fluorescence light that are independent of $m$.
	On the other hand, 
	the coherences of the initial state density matrix will play a role
	for scenarios with a spatial inhomogeneity other than the twisted beam.
	For instance the angular distributions do depend on $m$
	for the collision of a twisted particle with an inhomogeneous
	target, like:
	a second beam of twisted particles \cite{Ivanov:PRD2011,Ivanov:PRD2012},
	a localized microscopic target such as a single
	atom \cite{Lloyd:PRL2012,Matula:JPB2013,VanBoxem:PRA2014,VanBoxem:PRA2015,Surzhykov:PRA2015},
	or a quantum dot \cite{Quinteiro:PRA2015}.

	Another possibility to recover the coherences in the off-diagonal
	elements of the twisted density matrix is to look for the angular
	momentum of the scattered particles.
	In fact, it has been shown in \cite{Jentschura:PRL2011,Jentschura:EPJC2011}
	that Compton backscattered photons do indeed carry orbital angular momentum.
	In order to access the angular momentum of the scattered photons,
	one needs to determine the final-state density operator $\op \rho_f$ in the the basis
	of the twisted states. This requires a suitable detection operator that
	directly measures the orbital angular momentum of the scattered photons
	\cite{book:Balashov,Ivanov:PRD2011,Jentschura:PRL2011}.

	\subsection{Angular distribution and Stokes parameters for the scattering of a twisted photon with well-defined TAM}
	
	If we substitute the initial state density matrix \eqref{eq:rho.twisted} of the twisted photon
	into Eq.~\eqref{eq:reduced.density.matrix.finals}, and by performing the integration over
	the plane-wave basis in cylindrical coordinates
	$\int \!\widetilde{\di \vec k} = \frac{V}{(2\pi)^3} \int \! \di \varphi_k \di k_z \di k_\perp k_\perp$,
 	we obtain the reduced density
	matrix for the Compton scattering of a twisted photon
	(in the dipole approximation $\omega_f=\omega_i$)	.
	It includes an integration over all plane-wave components as described by $\varphi_k$
	\begin{align}
		\rho_{\Lambda' \Lambda} =
			 \elr^2
			\int\!\frac{\di\varphi_k}{2\pi}
			\left(
				\vec{\varepsilon}_{\vec{k}_f\Lambda'}^* \cdot
				\vec{\varepsilon}_{\vec{k}(\varphi_k)\Lambda_i}
			\right)^*
			\left(
				\vec{\varepsilon}_{\vec{k}_f\Lambda}^* \cdot
				\vec{\varepsilon}_{\vec{k}(\varphi_k)\Lambda_i}
			\right) \,,
	\label{eq:reduced.density.matrix.tw}
	\end{align}
	with $\vec k(\varphi_k)$ from Eq.~\eqref{eq:k.cone}.
	This reduced density matrix
	can be directly	compared with the corresponding result for plane-wave light,
	Eq.~\eqref{eq:reduced.density.pw}.
			
	\begin{figure}[!th]
		\includegraphics[width=\columnwidth]{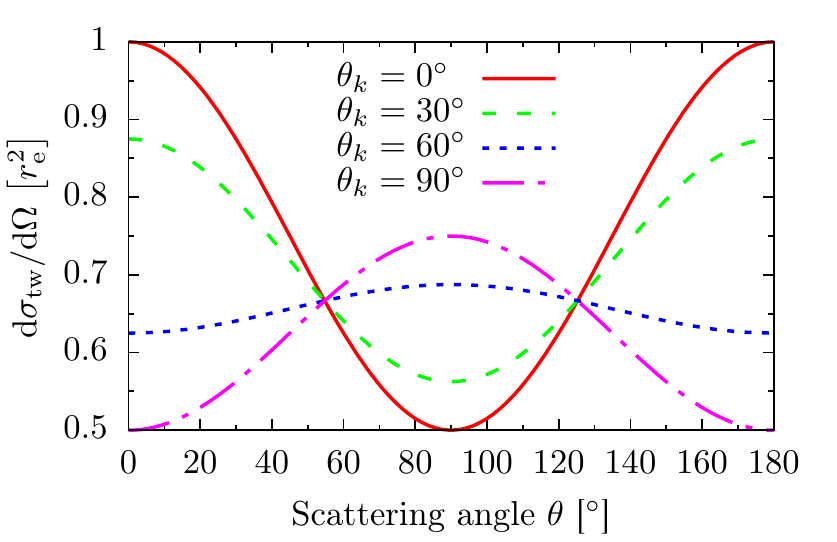}
		\caption{
			(Color online) 
			Differential cross section $\di\sigma_\mathrm{tw}/\di\Omega$ for Compton scattering of twisted
			light as a function of the scattering angle $\theta$.
			Results are shown for various values of the momentum cone
			opening angle $\theta_k$ of the twisted photons.
			The case $\theta_k=0$ is equivalent to plane-wave photons.
			Of course, the limit $\theta_k \to \unit{90}{\degree}$ of large
			cone opening angles can not be reached experimentally.		
		}
		\label{fig:tw-cs}
	\end{figure}

	If we make use of the explicit representation of the polarization vectors \eqref{eq:def.polvector},
	the integration over the azimuthal angle $\varphi_k$
	in Eq.~\eqref{eq:reduced.density.matrix.tw} can be carried out and
	yields the differential cross section
	\begin{align}
		\frac{\di\sigma_\mathrm{tw}}{\di\Omega} =  \frac{\elr^2}{4} 
			\left[ 
				(1+\cos^2\theta_k)(1+\cos^2\theta) + 2\sin^2\theta_k\sin^2\theta
			\right] \,,
		\label{eq:cs-tw}
	\end{align}
	where $\theta$ denotes the scattering angle measured from the $z$-axis,
	and $\theta_k = \arctan \kappa_\perp/\kappa_\parallel$ denotes the opening angle of the
	initial twisted photon beam.
	As anticipated above, the differential cross section is independent of the
	value of the projection of total angular momentum $m$, but
	it does depend on the momentum cone opening angle $\theta_k$. Note that in
	the limit $\theta_k\to 0$ we recover the plane-wave result \eqref{eq:pw-cs}.
	
	The results for the differential cross section is shown in Fig.~\ref{fig:tw-cs} as function of the 
	scattering angle $\theta$ for various cone angles $\theta_k$.
	The red solid curve ($\theta_k=0$) corresponds to the case of initial plane-wave photons and
	shows the well known symmetric angular distribution which is minimal at the scattering angle
	$\theta=\unit{90}{\degree}$, where
	it is just $1/2$ of the value at forward or backward scattering.
	For increasing values of the cone opening angle $\theta_k>0$ the angular distribution
	of the scattered photons gradually changes and the dip at $\theta=\unit{90}{\degree}$
	becomes less pronounced.
	For sufficiently large $\theta_k>\theta_k^\star$ the distribution
	turns around with the maximum of the angular distribution occurring at $\theta=\unit{90}{\degree}$
	(e.g.~for the blue dotted curve in Fig.~\ref{fig:tw-cs}).
	This crossover occurs at the ``magic angle'' of
	$\theta_k^\star = \arccos(1/\sqrt{3})\approx\unit{54.7}{\degree}$, where
	the angular distribution is flat.

		\begin{figure*}[!th]
		\begin{tabular}{@{}c@{}c@{}c@{}}
			\footnotesize(a)~Stokes Parameter $P_1$&%
			\footnotesize(b)~Stokes Parameter $P_3$&%
			\footnotesize(c)~Degree of polarization $\Pi$\\%
			\includegraphics{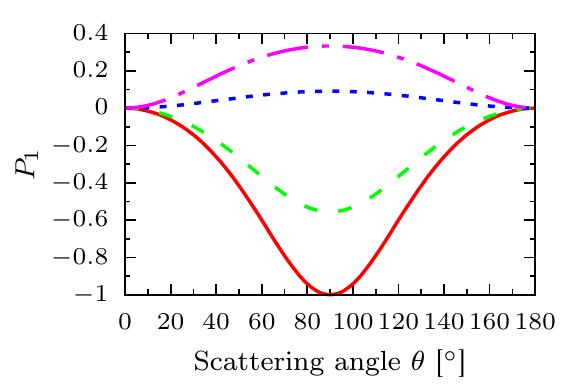}&%
			\includegraphics{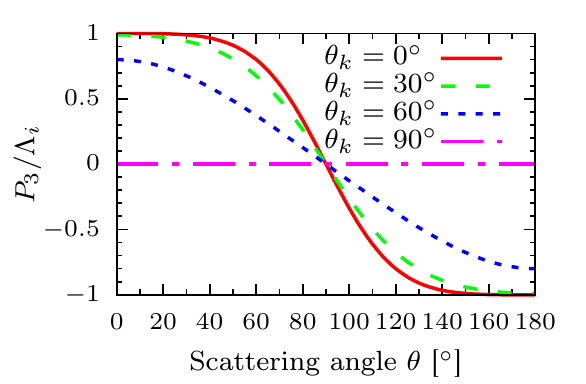}&%
			\includegraphics{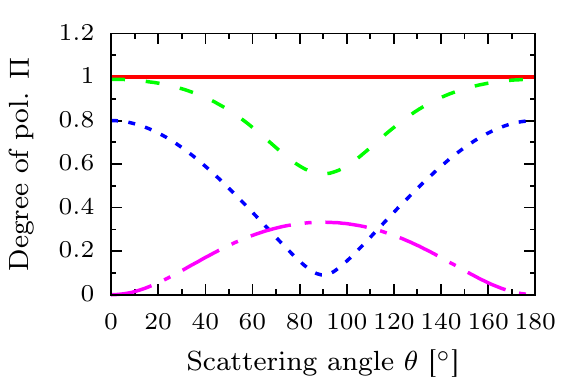}
		\end{tabular}
		\caption{
			(Color online) 
			The polarization of Compton scattered photons is characterized by the			
			Stokes parameters $P_1$ (left), $P_3$ (middle) and the degree of 
			polarization $\dop$ (right) as a function of the scattering angle.
		Results are shown for beams of
		 twisted photons with different momentum cone opening angles $\theta_k$.
		As in Fig.~\ref{fig:tw-cs}, the case $\theta_k=0$ is equivalent to plane-wave photons.
		}
		\label{fig:tw-stokes}
	\end{figure*}	

	In addition to the angular distribution discussed above, the momentum cone opening angle $\theta_k$ of the
	twisted photon beam also influences the polarization properties of the Compton scattered light.	
	For incident photons in the Bessel state,	
	the Stokes parameters are given by
	\begin{align}
		P_1
		&= \frac{(1-3\cos^2\theta_k)\sin^2\theta}
			{(1+\cos^2\theta_k)(1+\cos^2\theta) + 2\sin^2\theta_k\sin^2\theta}
			\,, \\
		P_2 &= 0\,, \\
		P_3
		&= \frac{4\Lambda_i\cos\theta_k\cos\theta}
			{(1+\cos^2\theta_k)(1+\cos^2\theta) + 2\sin^2\theta_k\sin^2\theta}
			\,.
			\label{eq:P3.tw}
	\end{align}
	Similar to the differential cross section,
	the Stokes parameters do neither depend on the azimuthal angle of the scattered photon,
	nor on the total angular momentum
	$m$ of the incident twisted beam,
	as the whole scenario is cylindrically symmetric.
	Again the results for the plane-wave case are reproduced for $\theta_k=0$.

	The results for the Stokes parameters are shown in Fig.~\ref{fig:tw-stokes} as function of the 
	scattering angle $\theta$ for different opening angles $\theta_k$.
	As discussed above for plane-wave photons, the values of $P_1$ and $P_3$,
	which quantify the linear and circular polarization of the scattered radiation, respectively,
	depend on the scattering angle.
	Both values are sensitive to the cone opening angle $\theta_k$ of the twisted light.
	For instance, as depicted in Fig.~\ref{fig:tw-stokes} (a),
	the value of $P_1$ at $\unit{90}{\degree}$ scattering decreases in magnitude
	for increasing values of $\theta_k$, starting from $P_1=-1$ for the plane-wave case ($\theta_k=0$).
	In particular, for the ``magic angle'' of
	$\theta_k^\star = \arccos(1/\sqrt{3})$ the scattered photons are
	not linearly polarized, since $P_1=0$ for all scattering angles.
	For even larger $\theta_k>\theta_k^\star$ the value of $P_1$ is positive which indicates
	a change of the plane of linear polarization of the scattered photons which are now
	(partially) polarized in the scattering plane.
	Since the sign of $P_3$ depends on the helicity 
	$\Lambda_i=\pm1$ of the incident photons, we show in Fig.~\ref{fig:tw-stokes} (b) the combination
	$P_3/\Lambda_i$ instead. 
	The values of $P_3$ gradually decrease for increasing $\theta_k$,
	and approach zero for $\theta_k \to \unit{90}{\degree}$.
	For twisted light with $\theta_k>0$, the degree of polarization is smaller than $1$, so the 
	scattered photons are not fully polarized anymore. For not too large
	cone opening angles $\theta_k$ the scattered radiation is depolarized the strongest
	at $\theta=\unit{90}{\degree}$. For sufficiently large $\theta_k$ angular dependence of the degree of
	polarization shows the opposite behavior. In particular for $\theta_k\to \unit{90}{\degree}$
	the scattered radiation becomes completely depolarized for forward and backward scattering,
	while the degree of polarization $\dop$ is nonzero at finite scattering angles, see Fig.~\ref{fig:tw-stokes} (c).

	\subsection{Angular distribution for the scattering of a superposition of twisted photons}
	
	As discussed above, the angular distribution of Compton scattered photons and their polarization
	does not depend on the value of the total angular momentum $m$ if the initial
	light is prepared in a Bessel state with well-defined $m$;
	it just depends on the opening angle $\theta_k$ of the beam.
		
	We now examine the Compton scattering of a coherent superposition of two states with
	equal longitudinal and transverse momenta, $\kappa_\perp$ and $\kappa_\parallel$,
	equal helicity $\Lambda$, but with two different 
	values of total angular momentum $m_2 > m_1$. 
	Such a superposition is described by the state vector
	\begin{align}
	\ket{\gamma} =
				c_1\,	\ket{\kappa_\perp \kappa_\parallel m_1 \Lambda} 
			+ c_2 \, \ket{\kappa_\perp \kappa_\parallel m_2 \Lambda} 
		\,,
		\label{eq:state.2tw}
	\end{align}
	where the coefficients fulfil $|c_1|^2+|c_2|^2=1$ and the state $\ket{\gamma}$
	is normalized $\scp{\gamma}{\gamma}=1$. 
	The state vector \eqref{eq:state.2tw} lies on the Bloch sphere with the two basis
	states as the poles because $\ket \gamma$ is a pure quantum state \cite{Yao:AdvOP2011}.
	The experimental generation of such superpositions of twisted
	states has been reported, e.g., in \cite{Vasilyeu:OptEx2009}.

	Such superpositions of twisted beams have been considered previously
	in theoretical studies of scattering and atomic absorption processes,
	e.g.~in Refs.~\cite{Ivanov:PRD2011,Seipt:PRA2014b,Surzhykov:PRA2015}.
	In these previous studies it was seen already that the interference 
	between these two TAM
	eigenstates leads to angular distributions of the scattered
	particles that depend on the difference $\Delta m= m_2-m_1$
	of the total angular momentum values of the two beams.
	This interference between the two components of the photon state $\ket{\gamma}$,
	Eq.~\eqref{eq:state.2tw}, can be seen in the off-diagonal elements of the
	density matrix of the initial photon state $\ket \gamma$ in the plane-wave basis
	\begin{multline}
	\exv{\vec k\lambda}{(\op\rho_\gamma)_{m'm}}{\vec k\lambda}
	 = \delta_{\lambda\Lambda} \frac{(2\pi)^2}{\kappa_\perp V}
	 \delta(k_\perp-\kappa_\perp) \delta(k_\parallel - \kappa_\parallel)\\
	\times
	\begin{pmatrix}
	|c_1|^2 & c_1 c_2^* \, e^{-i\Delta m (\varphi_k-\pi/2)} \\
	c_1^* c_2 \, e^{i\Delta m (\varphi_k-\pi/2)}			& |c_2|^2 
	\end{pmatrix}
	\label{eq:density.2tw}
	\end{multline}	
	The interferences are
	maximized by choosing coefficients $c_n$	
	with equal modulus and a relative phase $\delta$ as, e.g.,
	$c_1=1/\sqrt{2}$ and $c_2 = e^{i\delta}/\sqrt{2}$.
	These equal-weighted superpositions are all lying on the equator of the Bloch sphere,
	and where the relative phase $\delta =0\ldots 2\pi$ denotes the longitude \cite{Yao:AdvOP2011}.

	Let us note here that still only the momentum-diagonal terms
	of the photonic density matrix $\op \rho_\gamma$
	do contribute to the scattering cross section.
	The superpostion of more than one twisted state modifies the distribution of
	plane-wave components on the momentum cone
	so that they are no longer uniformly distributed on the cone as for the
	case of a single Bessel state.
	Instead, the azimuthal distribution
	of plane wave modes is modulated by the difference of the
	total angular momentum of the two beams $\Delta m$. 
	This is described by the off-diagonal elements in the density matrix \eqref{eq:density.2tw}
	in the space of the two basis states of the superposition \eqref{eq:state.2tw}.

	For such a superposition we readily obtain the differential cross section in the dipole approximation
	\begin{multline}
		\frac{\di\sigma_\mathrm{2tw}}{\di\Omega} 
		= \elr^2 
			\sum\limits_{\Lambda_f} 
			\int\!\frac{\di\varphi_k}{2\pi}
			\left[
				1 + \cos\left( 
					\Delta m \left(\varphi_k-\tfrac{\pi}{2}\right) + \delta 
				\right)
			\right] \\
		\times
			\left|
				\vec{\varepsilon}_{\Lambda_f}^*(\theta,\varphi)\cdot
				\vec{\varepsilon}_{\Lambda_i}\left(\theta_k,\varphi_k\right)
			\right|^2 \,,
	\end{multline}
	which can be split into two terms, 
	\begin{align}
		\frac{\di\sigma_\mathrm{2tw}}{\di\Omega} 
		= \frac{\di\sigma_\mathrm{tw}}{\di\Omega} 
		+ \frac{\di\sigma_\mathrm{int}}{\di\Omega}\,.
	\end{align}
	The first term, with the ``$1$'' in the square brackets,
	is the cross section for the case of a single TAM eigenstate, already discussed in the
	previous section, 
	cf.\@ Eq.~\eqref{eq:cs-tw}.
	The second term $\di\sigma_\mathrm{int}/\di\Omega$
	describes the interference between the two superimposed TAM eigenstates,
	and is related to the off-diagonal elements of the density matrix in Eq.~\eqref{eq:density.2tw}
	by means of
	$\cos\left( 
					\Delta m \left(\varphi_k-\tfrac{\pi}{2}\right) + \delta 
				\right)$,
	which describes the azimuthal modulation of the density of plane-wave states on the momentum cone.

	By performing the integration over the azimuthal angle $\varphi_k$ the interference
	of these eigenstates
	contributes to the cross section
	\begin{align}
		\frac{\di\sigma_\mathrm{int}}{\di\Omega} = 
		\begin{cases}
			\frac{\elr^2}{8} \sin\left(\varphi+\delta\right) 
				\sin(2\theta_k)\sin(2\theta)\,, 
			& \Delta m = 1\,, \\
			-\frac{\elr^2}{8} \cos\left(2\varphi+\delta\right) 
				\sin^2\theta_k\sin^2\theta\,,
			& \Delta m = 2\,, \\
			0\,, 
			& \Delta m \geq 3\,. \label{eq:int-cs-circ}
		\end{cases}
	\end{align}
	if $\Delta m=1$ or $2$ and vanishes otherwise
	due to dipole selection rules \cite{Surzhykov:PRA2015}.
	When we keep the first order non-dipole correction due to the electron recoil in the
	frequency of the scattered photons $\omega_f \neq \omega_i$, Eq.~\eqref{eq:omegaf},
	we find a non-vanishing interference contribution also for $\Delta m=3$, 
	\begin{align}
		\frac{\di\sigma_\mathrm{int}}{\di\Omega} = -
		\frac{\hbar\omega_i}{\elm c^2}
		\frac{\elr^2}{16}
		\sin( 3\varphi+\delta)\sin^3\theta_k\sin^3\theta\,,
	\end{align}
	which is proportional to the small recoil parameter 
	$\hbar \omega_i/\elm c^2\ll1$,
	and is therefore much smaller than the interference terms for $\Delta m= 1,2$.
	
		\begin{figure*}[!th]
		\begin{tabular}{m{1em}|cccc}
			& $\theta_k=\unit{0}{\degree}$ & $\theta_k=\unit{30}{\degree}$ & $\theta_k=\unit{60}{\degree}$ 
			& $\theta_k=\unit{90}{\degree}$ \\\hline
			\rotatebox{90}{$\Delta m=1$} &
			\noindent\parbox[c]{0.2\hsize}{\includegraphics{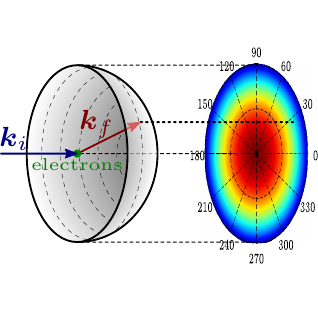}}&
			\noindent\parbox[c]{0.2\hsize}{\includegraphics{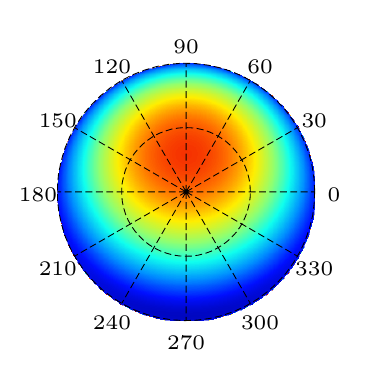}}&
			\noindent\parbox[c]{0.2\hsize}{\includegraphics{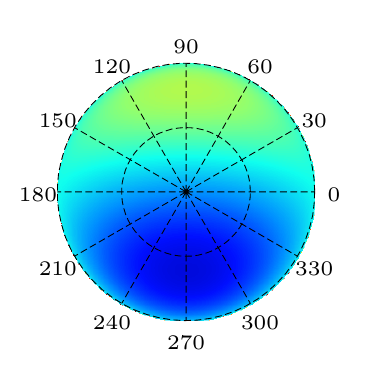}}&
			\noindent\parbox[c]{0.2\hsize}{\includegraphics{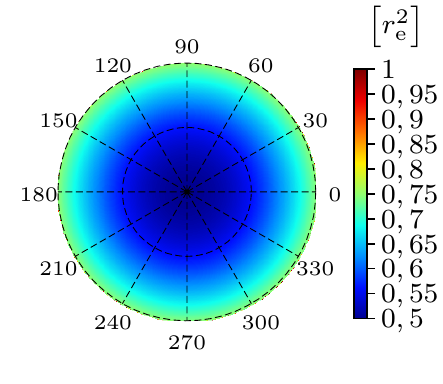}}\\[-0.5em]
			\rotatebox{90}{$\Delta m=2$} &
			\noindent\parbox[c]{0.2\hsize}{\includegraphics{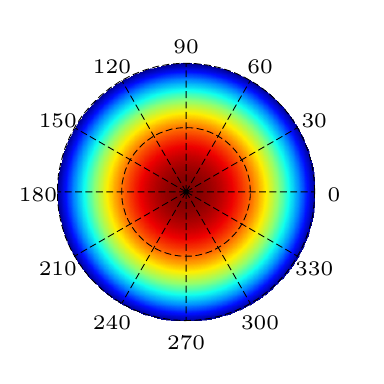}}&
			\noindent\parbox[c]{0.2\hsize}{\includegraphics{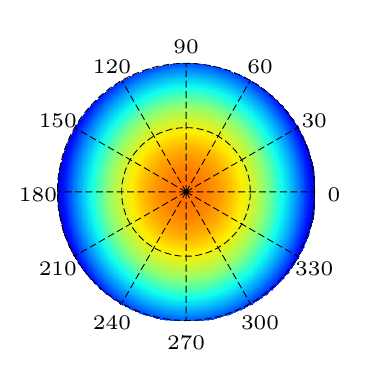}}&
			\noindent\parbox[c]{0.2\hsize}{\includegraphics{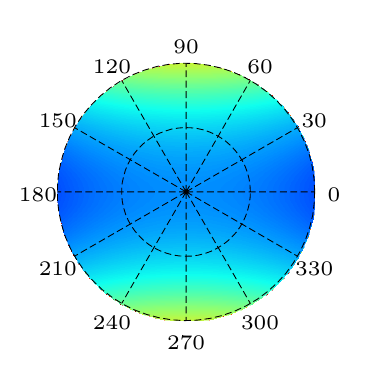}}&
			\noindent\parbox[c]{0.2\hsize}{\includegraphics{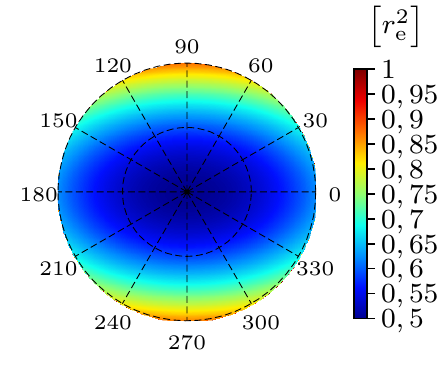}}
		\end{tabular}
		\caption{
			(Color online) Angular differential cross section $\di\sigma_\mathrm{2tw}/\di\Omega$
			of Compton scattered light for initial photons in a superposition of twisted waves with
			$\Delta m = 1$ (top) and $\Delta m=2$ (bottom)
			for different cone opening angles $\theta_k$, and for $\delta=0$.	
			The disks represent a
			projection of the forward scattering hemisphere (see the drawing in the top left panel), i.e.
			the radial axis of the disks represents $\sin\theta$
			for scattering angles $0\leq\theta\leq\unit{90}{\degree}$, and
			the azimuthal axis represents the 
			azimuthal angle $\varphi$ of the scattered photon.
			For $\theta_k=0$ (left column), again, the cross section
			coincides with the results of an incident plane-wave photon;
			they are equal for $\Delta m=1$ and $\Delta m=2$ and they are independent of
			the azimuthal angle $\varphi$.
		}
		\label{fig:2tw1}
	\end{figure*}

	Figure \ref{fig:2tw1} displays	
	the angular distributions of Compton scattered photons
	for an incident superposition
	of two twisted Bessel photons
	for $\Delta m=1,2$ and for different cone opening angles $\theta_k$.
	Obviously, the interference term depends on the azimuthal angle 
	$\varphi$ of the scattered photon, and not on the scattering angle 
	$\theta$ alone.
	The disks in Fig.~\ref{fig:2tw1} represent a projection of the forward hemisphere of scattered photons,
	as drawn in the top left panel.
	The superpositions of twisted photons break the axial symmetry of the initial state.
	The number of azimuthal modulations of plane-wave states on the momentum cone
	is just $\Delta m$ and the orientation of that pattern is determined by the relative phase $\delta$,
	i.e.~by the latitude of the state $\ket \gamma$ on the Bloch sphere.
	Changing the value of $\delta$ results in a rotation of the distributions
	of scattered photons in Fig.~\ref{fig:2tw1} with respect to the azimuthal angle $\varphi$.

	Let us briefly discuss the angular distribution in the backward-hemisphere, i.e.~for scattering
	angles $\theta \geq \unit{90}{\degree}$. The distributions can be easily read off
	Eqs.~\eqref{eq:cs-tw} and \eqref{eq:int-cs-circ}.
	In particular, for the case $\Delta m=1$ we obtain the relation
	$\di\sigma_\mathrm{2tw}^{(\Delta m = 1)}/\di\Omega(\theta,\varphi)
		=
		\di\sigma_\mathrm{2tw}^{(\Delta m = 1)}/\di\Omega(\pi - \theta, - \varphi-2\delta)$.
	For $\Delta m=2$, on the other hand, we obtain the symmetry relation
	$\di\sigma_\mathrm{2tw}^{(\Delta m = 2)}/\di\Omega(\theta,\varphi)
		=
		\di\sigma_\mathrm{2tw}^{(\Delta m = 2)}/\di\Omega(\pi - \theta,\varphi)$.

	\section{Conclusions}\label{sec:conclusions}
	
	In summary, a theoretical study has been performed for the Compton scattered of photons from a
	Bessel beam on electrons in the rest frame of the electrons.
	In the long-wavelength limit of the incident radiation
	and based on the non-relativistic Schr\"o{}dinger's equation,
	the density matrix theory has been used to
	analyze the angle-differential cross section as function of the scattering angle and for Bessel
	beams with different cone opening angles and total angular momentum.
	
	Our formulation of the problem in the density matrix formulation explains
	why the angular distribution and polarization of the scattered photons
	do not depend on the value of the projection of angular momentum:
	Because of the symmetry of the system only
	the those elements of the incident twisted photon density matrix contribute
	to the reduced density matrix of the scattered photons that are diagonal in the
	momentum quantum numbers.
	The angular momentum value appears only via the difference of the vortex phase factors
	of the plane-wave components of the twisted photon
	and, therefore, vanishes on the diagonal of the density matrix.
	
	We found completely analytical results for the angular and polarization distribution
	of Compton scattered photons for the scattering of twisted photons.
	These distributions are sensitive with regard to the 
	momentum cone opening angle $\theta_k$.
	In particular we observed a depolarization of the
	scattered radiation for large values of this cone opening angle.
	In addition, it was found that the angular distributions of the scattered photons for a
	superpositions of twisted photon beams with different $m$
	differ from the case of a single TAM eigenstate
	if $\Delta m=1$ or $2$ due to dipole selection rules.
	These differences vanish for $|\Delta m|>2$ in the dipole approximation and
	remain small if non-dipole contributions are taken into account.

	For beams of photons and/or electrons of higher energy,
	a relativistic treatment of the electron-photon interaction might be more appropriate
	instead but should not change the central results of this work as long as the changes in
	the recoil parameter are moderate.
	In fact, as long as the recoil parameter $r$ in Eq.~\eqref{eq:recoil} is small,
	our results can be directly translated to scenarios of inverse Compton scattering
	by applying a suitable Lorentz transformation.
	For inverse Compton scattering, low-frequency photons (e.g.~optial laser photons)
	are scattered on ultra-relativistic electrons, and the backscattered photons' frequency
	is Doppler up-shifted to the x-ray regime.

	
	\appendix*

	\section{Normalization of quantum states of plane-wave and twisted photons}	
	\label{sec:normalization}

	In this Appendix we discuss all necessary details
	on the normalization of the quantum states that enter the calculation
	of the final-state density matrix in the non-relativistic Compton scattering
	of plane-wave or twisted light.
	The usual normalization in a finite box does not work in this case,
	because the twisted photons are cylindrical symmetric modes.
	Moreover, the states of twisted photons
	are represented as a \textit{continuous}
	coherent superposition of plane waves,
	while in a finite-sized box the momentum modes are discrete.
	Therefore, we need to quantize the modes
	in an \textit{infinite} volume $V \to \infty$.
	In order to correctly normalize twisted-particle quantum states we
	will explicitly keep all the factors of the formally infinite volume.
	We normalize \textit{all} one-particle quantum states $\ket{\psi}$, i.e.~for the
	electrons and the plane-wave or twisted photons, to unity: $\scp{\psi}{\psi}=1$.
	In the following we discuss what this implies in detail for the electron and photon states,
	their density operators, spatial wavefunctions, as well as for the spatial probability density.

	\subsection{Plane-wave electron states}

	Throughout our paper, all initial and final states of the electrons are described as plane waves, i.e.~as
	momentum eigenstates $\vop{p} \ket{p} = \vec{p} \ket{p}$,
	where $\vop p$ is the electron momentum operator, and the states are characterized by
	the three quantum numbers of the linear momentum eigenvalue $\vec p = (p_x,p_y,p_z)$.
	We require that the one-particle states are normalized as
	\begin{align}
	\scp{\vec p}{\vec p}=1 \,,
	\label{eq:app.norm.el}
	\end{align}
	and, hence, for the orthogonality relation
	\begin{align}
	\scp{\vec p'}{\vec p} = \left(\frac{2\pi}{L} \right)^3 \delta( \vec p' -\vec p) \,,
	\label{eq:app.ortho.el}
	\end{align}
	and where the normalization \eqref{eq:app.norm.el} follows from \eqref{eq:app.ortho.el} with the usual
	interpretation of $\lim_{\vec p'\to \vec p} \delta(\vec p' -\vec p) = (L/2\pi)^3$ \cite{book:Peskin}.

	We may define a \textit{regularized} delta distribution
	\begin{align}
	\widetilde \delta( \vec p' -\vec p ) := \left(\frac{2\pi}{L} \right)^3 \delta( \vec p' -\vec p)\,,
	\end{align}		
	which has the property $\lim_{\vec p'\to \vec p}  \widetilde \delta(\vec p' - \vec p ) = 1$.
	The properly normalized integration measure for these plane-wave states is given by
	\begin{align}
	\int \! \widetilde{\di \vec p} :=  \int \! \left(\frac{L}{2\pi}\right)^3\di^3 \vec p \,.
	\label{eq:measure.el}
	\end{align}
	In particular, $\int \! \widetilde{\di \vec p} \, \widetilde{\delta} (\vec p-\vec p') = 1$.
	These definitions also provide the correct way of counting the number of final states; their density
	is just $\widetilde{\di \vec p}$. 
	Moreover, the completeness relation for the plane-wave electron states is
	\begin{align}
	 \int \! \widetilde{\di \vec p} \, \dya{\vec p}{\vec p} = \op 1\,.
	\end{align}		
	Because the states $\ket{\vec p}$ are normalized, the trace of the density operator
	$ \op \rho_{\vec p}= \dya{\vec p}{\vec p}$ is just unity ${\rm tr} \, ( \op \rho_{\vec{p}} ) = 1$,
	where we also have to take the integration measure \eqref{eq:measure.el} when
	calculating the trace of the density matrix.

	 The position space wave function is just given by
	 $\psi_{\vec{p}}( \vec x) = \scp{\vec x}{\vec p} =e^{i \vec p \cdot \vec x} / \sqrt{V}$.
	The particle density is determined as the expectation value of the
	particle density operator
	$\op n(\vec x) = \dya{\vec x}{\vec x}$, as $n_{\vec p}(\vec x) = {\rm tr} \, \op n(\vec x) \op \rho_{\vec p}$
	and just yields a constant local particle density of $n_{\vec p}(\vec x)  = 1/V$.
	Therefore one usually says that the plane-waves \eqref{eq:app.norm.el}
	are normalized to one particle in the infinite quantization volume $V=L^3\to \infty$.

	\subsection{Plane-wave photon states}

	We apply the same normalization to the plane-photon states as we employed
	for the plane-wave electron states in the previous subsection.
	The only small difference is that the photon states are characterized
	also by their helicity $\Lambda$ in addition to the linear momentum
	eigenvalue $\vec k$. We thus orthonormalize the plane-wave one
	particle photon states $\ket{\vec k \Lambda}$ as
	\begin{align}
	 \scp{\vec k\Lambda}{\vec k'\Lambda'}  =	\delta_{\Lambda\Lambda'} \widetilde{\delta} (\vec k - \vec k') \,.
	 \label{eq:app.norm.pw.phot}
	 \end{align}
	Moreover, we quantize the photon field, expanding the photon field operator
	\begin{align}
		\vop{A}(\vec{x}) 
		= \sumint \limits_\Lambda \! \widetilde{\di \vec{k}}  \, N_k
			\left[
				\op{c}_{\vec k \Lambda} \vec u_{\vec k \Lambda}(\vec x)
				+ 
				\op{c}_{ \vec k \Lambda}^\dagger \vec u_{\vec k \Lambda}^* (\vec x)
			\right] \,,
		\label{eq:vector-potential.app}
	\end{align}
	into a circularly polarized plane-wave basis
	$\vec u_{\vec k \Lambda}(\vec x) 
		= 
		e^{i\vec k\cdot \vec x} \, \vec{\varepsilon}_{\vec k \Lambda}$,
	and with the same meaning of $\widetilde{\di \vec k}$ as for plane-wave electrons.
	The normalization factor $N_k = \sqrt{ 2\pi c / k V }$
	is chosen in such a way that the free Hamiltonian of
	the photon field is given by
	\begin{multline}
	H_\gamma
	= \frac{1}{8\pi} \int \! \di^3 \vec x \, ( \vop E^2 + \vop B^2 )	\\
	\stackrel{!}{=} \sumint \limits_\Lambda \! \widetilde{\di \vec k} \,  \omega_k
	\left( 
		\op c_{\vec k\Lambda}^\dagger \op c_{\vec k\Lambda} + \frac{1}{2} 
	\right)\,,
	\end{multline}		
	where $\vop E$ and $\vec B$ are the electric and magnetic field operators, respectively.

	The one-particle states are generated by the photon creation operators
	from the vacuum (zero-photon) state
	\begin{align}
	\ket{\vec k\Lambda}  = \op c_{\vec k\Lambda}^\dagger \ket{0} \,.
	\end{align}
	The commutation relations of the photon creation and annihilation operators
	\begin{align}
	[\op c_{\vec k\Lambda} , \op c_{\vec k'\Lambda'}^\dagger]
	= \delta_{\Lambda\Lambda'} \widetilde{\delta} (\vec k - \vec k')
	\end{align}
	completely fix the orthonormalization \eqref{eq:app.norm.pw.phot}.
	 
	The proper measure of the final states now includes a sum over the helicity states, thus,
	the completeness relation is given by
	\begin{align}
	\sumint \limits_\Lambda \! \widetilde{\di \vec k}\,
	 \dya{\vec k\Lambda}{\vec k\Lambda} 
	 = 
	 \op 1 \,.
	\end{align}
	That means we need to include the sum over the two helicity states in the trace of the
	photonic density matrix in order have ${\rm tr}\, ( \op \rho_{\vec k\Lambda} )=1$, and
	with the same interpretation of having one particle per volume $V$ as above.
	The vector potential that corresponds to the plane-wave one-photon state can be calculated as
	\begin{multline}
	\vec A_{\vec k \Lambda}(\vec x) := \exv{0}{\vop{A}(\vec x)}{\vec k \Lambda} =
	N_k \vec u_{\vec k \Lambda} \\
	= \sqrt{\frac{2\pi c}{k V}} \vec{\varepsilon}_{\vec k \Lambda} e^{i\vec k\cdot \vec x}\,.
	\end{multline}

	\subsection{Twisted-wave photon states}
	
	Twisted photons are cylindrical symmetric modes and therefore need to be normalized to
	a cylindrical volume $V=\pi R^2L_z$ with radius $R$ and 
	height $L_z$ along the $z$-axis.
	The twisted photon states that we defined in Eq.~\eqref{eq:tw.superposition}
	are orthonormalized in the following sense:
	\begin{multline}
	\scp{\kappa_\perp' \kappa_\parallel' m' \Lambda'}{\kappa_\perp \kappa_\parallel m \Lambda} \\
	 = \frac{2\pi^2}{R L_z} \delta(\kappa_\parallel'-\kappa_\parallel) \delta(\kappa_\perp'-\kappa_\perp) \delta_{m'm} \delta_{\Lambda'\Lambda} \,,
	\end{multline}		
	which implies that
	$\scp{\kappa_\perp \kappa_\parallel m \Lambda}{\kappa_\perp \kappa_\parallel m \Lambda}=1$.
	To prove this normalization we make use of the 
	 identity for the radial delta function
	\begin{align}
	\lim_{\kappa_\perp'\to \kappa_\perp} \delta(\kappa_\perp'- \kappa_\perp) = \frac{R}{\pi} \,,
	\end{align}		
	where $R$ is the (infinite) radius of the cylindrical normalization volume.
	The above identification was proven, e.g., in Refs.~\cite{Jentschura:EPJC2011,Ivanov:PRD2011}.
	Moreover, for the longitudinal momentum delta function we employ the usual relation
	$\delta(\kappa_\parallel = 0) = L_z/2\pi$, where $L_z$ is the height of the quantization cylinder.

	The probability density that can be attributed to the twisted one-particle states
	$n_{\kappa_\perp \kappa_\parallel m \Lambda}(\vec x) 
	= {\rm tr}\, ( \op n(\vec x) \op \rho_{\kappa_\perp \kappa_\parallel m \Lambda})$
	turns out the be not spatially constant,
	but instead is given by
	\begin{align}
	n_{\kappa_\perp \kappa_\parallel m \Lambda}(\vec x) 
	&=  \frac{\kappa_\perp}{2L_z R} J_m^2(\kappa_\perp x_\perp) \,,
	\label{eq:local.density.tw}
	\end{align}		
	where $J_m$ are the Bessel functions of the first kind \cite{book:Watson}.

	The spatially averaged probability density
	\begin{align}
	\mean{n} := \frac{1}{V}  \! \int \! \di^3 \vec x \, n(\vec x) = \frac{1}{V} \, {\rm tr} \op \rho
	\end{align}
	which enters the definition of the cross section, Eq.~\eqref{eq:cross.section},
	is just proportional to the trace of the density operator because the position eigenstates
	form a complete basis.
	That means, also for twisted photons a normalized density
	operator correspond to one particle per volume $V=\pi L_z R^2$.
	The spatially averaged probability density can also be calculated directly from
	Eq.~\eqref{eq:local.density.tw} as
	\begin{align}
	 \mean{n_{\kappa_\perp \kappa_\parallel m \Lambda}}
	 &=  \frac{1}{V} \! \int \! \di\varphi \, \di z \, \di x_\perp \, x_\perp   
				 \frac{\kappa_\perp }{2L_z R} J_m^2(\kappa_\perp x_\perp) \nonumber \\
	 &= \frac{1}{V} \frac{\pi }{ R} 
	 \! \int \limits_0^R \!  \di x_\perp \,
	 		 x_\perp \kappa_\perp  \,  J_m^2(\kappa_\perp x_\perp) \,,
	\end{align}		
	and by approximating the radial integral over $x_\perp$ for large $R\to \infty$ by using the
	asymptotic expansion of the Bessel function for large arguments,
	$J_m(x) \approx \sqrt{2/\pi x } \, \cos(x- m\pi/2-\pi/4)$
	\cite{book:Watson,Jentschura:EPJC2011}, yielding
	\begin{align}
		\int\limits_0^R \di x_\perp \,  \kappa_\perp x_\perp \, J_m^2(\kappa_\perp x_\perp) 
		\approx \frac{R}{\pi} \,.
	\end{align}

	The vector potential that corresponds to the twisted one-photon states is given by	
	\begin{multline}
	\vec A_{\kappa_\perp \kappa_\parallel m \Lambda }(\vec x)
			= \exv{ 0 }{\vop A(\vec x)}{ \kappa_\perp \kappa_\parallel m \Lambda } \\
			=\sqrt{\frac{2\pi^2c^2}{\omega L_z R}} 
			 \int \! \frac{\di^2 \vec k_\perp}{(2\pi)^2} a_{\kappa_\perp m}(\vec k_\perp)
			\vec{\varepsilon}_{\vec k \Lambda} e^{i\vec k_\perp \cdot \vec x_\perp + i \kappa_\parallel z} 
			\,.
	\end{multline}
	Except for the different normalization factor in front of the integral, this coincides with the vector potential
	employed in Ref.~\cite{Matula:JPB2013} to define the twisted light.
	It was argued in \cite{Matula:JPB2013} that $A^\mu_{\kappa_\perp \kappa_\parallel m \Lambda}(\vec x)$
	describes a beam with well-defined projection of total angular momentum $m$.
	The product of the amplitude $a_{\kappa_\perp m}(\vec k_\perp)$ and the plane-wave
	polarization vector $\vec{\varepsilon}_{\vec k \Lambda}$ was shown to be an eigenfunction of the
	$z$-component of the total angular momentum operator $\op J_z$ with the eigenvalue $m$.

	Performing the momentum integrations in the above Fourier integral we obtain for the twisted-wave
	vector potential
	\begin{multline}
	\vec A_{\kappa_\perp \kappa_\parallel m \Lambda }(\vec x) 
		= \sqrt{\frac{2\pi^2c^2}{\omega L_z R}}  	\sqrt{\frac{\kappa_\perp}{2\pi}} \, e^{i \kappa_\parallel z} \\
	\times		
		\sum_{m=0,\pm1} 
		(-1)^{m_s} c_{m_s}(\Lambda) J_{m-m_s}(\kappa_\perp x_\perp)
			e^{i(m-m_s)\varphi} \vec \eta_{m_s}
	\end{multline}
	where the sum over $m_s$ runs over all possible projections of the photon spin angular momentum
	onto the $z$-direction and accounts for the coupling of (the projections of) orbital angular momentum ($m_\ell$)
	and spin angular momentum ($m_s$) to the total angular momentum $m = m_\ell+m_s$.
	This representation of the twisted wave vector potential
	employs the unit vectors $\vec \eta_0 = (0,0,1)^T$ and
	$\vec \eta_{\pm1} = (1,\pm i,0)^T/\sqrt{2}$, and the coefficients
	$c_0=- \frac{1}{\sqrt2}\sin \theta_k $ and $c_{\pm1}=\frac{1}{2}(1\pm \Lambda \cos \theta_k)$
	with the momentum cone opening angle $\tan \theta_k = \kappa_\perp/\kappa_\parallel$.


%

\end{document}